\documentclass{article}
\usepackage{amsmath,amssymb}
\usepackage{frascatiphys}
\usepackage{graphicx}
\begin{document}
\title{A force beyond the Standard Model\\ --\\
 Status of the quest for hidden photons}
\author{
Joerg Jaeckel       \\[0.1cm]
{\em Institut f\"ur theoretische Physik, Universit\"at Heidelberg,} \\{\em Philosophenweg 16, 69120 Heidelberg, Germany} \\
}
\maketitle
\baselineskip=11.6pt
\begin{abstract}
In this note we discuss the search for new gauge forces beyond the Standard Model. In particular we give an overview for the simplest case of a new U(1), kinetically mixed with the Standard Model photon (hypercharge boson), a so-called hidden photon (also known as dark photon, heavy photon or $A^{\prime}$).  
\end{abstract}
\baselineskip=14pt

\section{Introduction}
The Standard Model (SM) features three gauge forces, electromagnetism, the weak force and the strong force. 
The corresponding gauge group is ${\rm U(1)}\times{\rm SU(2)}\times{\rm SU(3)}$. This is certainly not an obvious structure, and it
is therefore only natural to ask if these are the only ones or, if there are additional (gauge) forces.

In this note we concentrate on the simplest case: an extra U(1)$_{X}$ kinetically mixed with the electromagnetic/hypercharge U(1) of the Standard Model. This corresponds to an extra photon-like particle, the hidden photon (HP).

Of course, when looking for a new extra force we immediately have to confront the question why we have not seen it.
This is directly linked to the properties of the new gauge bosons responsible for the force.
In principle there are two ways in which particles can hide from observation in experiments. The first, and so far the most commonly investigated, is that the new particles are very heavy. In this case one needs a lot of energy to create them and forces mediated by them are of extremely short range. Moreover, heavy particles that are not protected by a new symmetry decay very fast. The second option is that the interactions between the new particles and those of the SM are extremely weak. In this case their effects would simply be too feeble to have been observed so far. This is why such particles are often referred to as belonging to a so-called Òhidden sectorÓ. Particles in these hidden sectors could even be very light, potentially even massless.

These two possibilities suggest that exploring new physics is in a sense (at least) two-dimensional. One needs to explore in the direction of higher mass and energy, as well as in the complementary direction of very weak couplings. This suggests two entirely different search strategies. Higher masses can be most directly explored in high energy experiments like, for example, the Large Hadron Collider (LHC) at CERN. Probing very weak couplings requires high precision/intensity/luminosity but can often be done at fairly low energies. 
Both approaches nicely complement each other as can be seen in the case of an extra U(1) that we discuss in this note.

\section{Hidden photons}\label{sec:hid}
Let us consider an extra U(1) gauge group. 
If all Standard Model particles are uncharged under this new gauge group the dominant interaction with ordinary matter is via kinetic 
mixing\cite{Holdom:1985ag} with the hypercharge U(1) gauge boson. This is encoded in the following Lagrangian,
\begin{eqnarray}
\label{LagKM}
\mathcal{L} \!\!&\supset&\!\! -\frac{1}{4} W^{a}_{\mu\nu}W^{a,\mu\nu}-\frac{1}{4} B_{\mu \nu} B^{\mu \nu}
- \frac{1}{4} X_{\mu \nu} X^{\mu \nu}
- \frac{\chi_{_Y}}{2} B_{\mu \nu} X^{\mu \nu}
\\\nonumber
\!\!&+&\!\! \frac{m_{X}^2}{2} X_{\mu} X^{\mu}
+\frac{1}{2}\frac{m^{2}_{W}}{g^2}(-gW^{3}_{\mu}+g^{\prime}B_{\mu})^{2}+\frac{1}{2}m^{2}_{W}(W^{1}_{\mu}W^{1,\mu}+W^{1}_{\mu}W^{1,\mu})
\\\nonumber
\!\!&+&\!\! \mbox{SM matter and Higgs terms},
\end{eqnarray}
where $B_{\mu}$ and $W_{\mu}$ denote the usual electroweak gauge fields and
$X_\mu$ denotes the hidden U(1) field with gauge coupling $g_{_X}$. Importantly the term $\frac{\chi_{_Y}}{2} B_{\mu \nu} X^{\mu \nu}$ introduces
a mixing between $X_\mu$ and $B_\mu$. 

The naive one loop estimate for the mixing parameter is
\begin{equation}
\chi_{_Y} \sim \frac{e g_{_X}}{6\pi^2}\log\left(\frac{m}{\Lambda}\right)
\end{equation}
where $m$ is the mass of a heavy particle coupled to both the new U(1) and hypercharge and $\Lambda$ is some cutoff scale.
In general models of field~\cite{Holdom:1985ag} and string 
theory~\cite{Dienes:1996zr,Lukas:1999nh,Abel:2003ue,Blumenhagen:2005ga,Abel:2006qt,Abel:2008ai,Goodsell:2009pi,Goodsell:2009xc,Goodsell:2010ie,Heckman:2010fh,Bullimore:2010aj,Cicoli:2011yh,Goodsell:2011wn} a wide range of 
kinetic mixing parameters are predicted, stretching from  $\chi\sim 10^{-12}$ to $\chi\sim10^{-3}$.

In general, the HP mass might result from a Higgs or a St\"uckelberg mechanism. In the first case a Higgs particle appears in the spectrum, with mass $\sim \sqrt{\lambda} m_{X}/g_X$ where $g_X$ is the hidden sector gauge coupling and $\lambda$ the Higgs self-coupling. Even if we take $g_X$ to be relatively small, the Higgs particle phenomenology tightly constrains this scenario, especially for sub-eV values of $m_{X}$\cite{Ahlers:2008qc}.  
However, for very small $g_{X}$, as one finds in large volume string scenarions\cite{Burgess:2008ri,Goodsell:2009xc,Cicoli:2011yh}, viable hidden Higgs models can be realized. The expected regions are shown inside the dotted lines in Fig.~\ref{prospects}. 
The St\"uckelberg case also occurs naturally in large volume string compactifications\cite{Goodsell:2009xc,Cicoli:2011yh}.
Typical expected parameter values are indicated by the dash-dotted lines of Fig.~\ref{prospects}.

At energies far below the electroweak scale and for small masses of the new gauge boson, $m_{X}\ll m_{W}$,  we can consider only the remaining light degrees of freedom 
and the mixing is directly with the photon,
\begin{equation}
\label{LagKM2}
\mathcal{L} \supset -\frac{1}{4} F_{\mu\nu}F^{\mu\nu}- \frac{1}{4} X_{\mu \nu} X^{\mu \nu}
- \frac{\chi}{2} F_{\mu \nu} X^{\mu \nu}+ \frac{m_{X}^2}{2} X_{\mu} X^{\mu} +j_{\mu}A^{\mu},
\end{equation}
where the mixing with the photon is related to that with the hypercharge via
\begin{equation}
\chi=\chi_{_{Y}}\cos(\theta_{W}),
\end{equation}
where $\theta_{W}$ is the Weinberg angle. For later convenience we have also included the coupling to the electromagnetic current $j_{\mu}$.

\begin{figure}[t]
    \begin{center}
         {\includegraphics[scale=0.55]{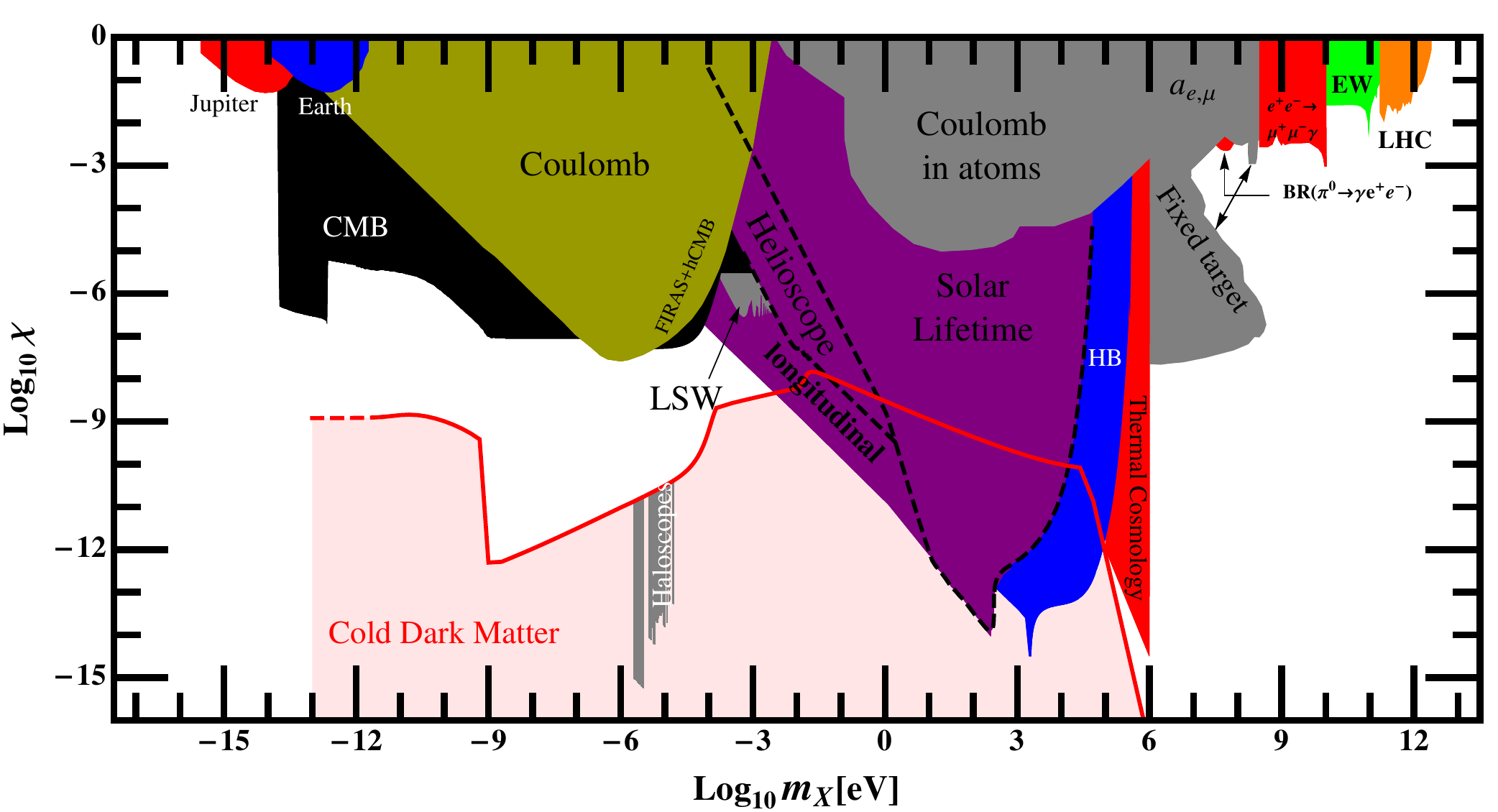}}
        \caption{\it Limits on the kinetic mixing of a hidden photon with the ordinary photon. Figure updated from\cite{Jaeckel:2010ni} with new and improved limits from\cite{Andreas:2012kx,Jaeckel:2012zr,Gninenko:2013qy,An:2013vn}. The area shaded in light red gives the area where HPs can be cold dark matter\cite{Arias:2012ly}.}
\label{exfig}
    \end{center}
\end{figure}

As we can see the kinetic mixing and the mass of the new particle are the only two new parameters. The current constraints are shown in Fig.~\ref{exfig}.
In the following sections we will sketch some of these constraints as well as prospects for future searches.

\section{A matter of convenience: a new force or photon--HP oscillations}

In Eq.~\eqref{LagKM2} we have introduced the somewhat unusual kinetic mixing term.
To get a better understanding it is convenient to remove this term by a suitable field re-definition.
There are two simple field re-definitions that we can use to remove the kinetic mixing term\footnote{Here and in the following we neglect terms of the order of $\chi^2$.}:
\begin{itemize}
\item[(1)]{} $A^{\mu}\rightarrow A^{\mu}-\chi X^{\mu}$.
\item[(2)]{} $X^{\mu}\rightarrow X^{\mu}-\chi A^{\mu}$.
\end{itemize}

Although the resulting physics is, of course, completely equivalent, the physical picture resulting from both shifts is somewhat different.
Depending on the situation it is often easier to use one or the other picture. Let us now briefly consider both pictures.

\subsection{Option (1): A $Z^{\prime}$ and a new force}
Inserting the shift $A^{\mu}\rightarrow A^{\mu}-\chi X^{\mu}$ into Eq.~\eqref{LagKM2} removes the kinetic mixing term,
\begin{equation}
\label{LagKM3}
\mathcal{L} \supset -\frac{1}{4} F_{\mu\nu}F^{\mu\nu}- \frac{1}{4} X_{\mu \nu} X^{\mu \nu}
+ \frac{m_{X}^2}{2} X_{\mu} X^{\mu} +j_{\mu}(A^{\mu}-\chi X^{\mu}).
\end{equation}

We now have two nicely independent particles. A massless particle $A$ and a massive particle $X$. $X$ is a massive uncharged vector particle. In that sense it is similar to a $Z$ boson and therefore it is a special case of a so-called $Z^{\prime}$.

A is the ordinary photon and couples to the electromagnetic current 
\begin{equation}
j^{\mu}=e n^{\mu},
\end{equation}
where $e$ is the electric charge and $n^{\mu}$ is the number current of charged particles with unit charge.

$X$ is the new hidden photon and it couples to ordinary matter via 
\begin{equation}
-\chi j_{\mu} X^{\mu}=-\chi e n_{\mu} X^{\mu}.
\end{equation}
In other words a particle with electromagnetic charge $Q$ now also carries a ``hidden charge'',
\begin{equation}
\label{xcharge}
Q_{X}=-\chi Q ,
\end{equation}
coupling it to the hidden photon.

It is now easy to calculate the interaction between two charged particles with charges $q_{1}e, q_{2}e$ separated by a distance $r$,
\begin{equation}
\label{coulomb}
V(r)=q_{1}q_{2}\frac{\alpha}{r}\left[1+\chi^2 \exp(-m_{X}r)\right].
\end{equation} 
The first term is, of course, the ordinary Coulomb interaction. The second is the additional contribution from the ``new force'' mediated by $X$.

Let us note that the form of the correction makes both ways of how to hide particles, mentioned in the introduction, explicit.
For a fixed resolvable distance $r_{\rm res}>0$ the new force becomes negligible whenever the mass $m_{X}\gg 1/r_{\rm res}$.
The force has a range shorter than the resolution and is therefore not observed.
Alternatively we can make $\chi$ very small. In this case the force becomes simply very weak and at some point unobservable even if $m_{X}$ is small
and the range is large.

Eq.~\eqref{coulomb} also provides a way to search for the new force: we can test Coulomb's law.
This can be done at scales of the order of $\sim\!\!10\,{\rm cm}$ with Cavendish experiments\cite{Williams:1971ms,Bartlett:1988yy,Popov:1999}.
The corresponding limit is shown in Fig.~\ref{exfig} labelled ``Coulomb'' and is currently the best limit in the range of $\mu{\rm eV}$. The precision achieved, about one part in $10^{16}$, is remarkable. Nevertheless it is also noteworthy that the best experiment is more than 40 years old.
At much smaller scales one can use atomic transitions\cite{Popov:1999,Karshenboim:2010ck,Karshenboim:2010cg,Jaeckel:2010xx} (see Fig.~\ref{exfig}). 
Following essentially the same arguments one can also look at 
magnetic fields, which is useful at much larger scales where one can use the magnetic fields of Earth and Jupiter which have been mapped with a good precision. This gives the bounds labelled ``Earth'' and ``Jupiter''\cite{Popov:1999}.

Other experiments do not look for the mediated force, but directly for the new exchange particle. 
From Eq.~\eqref{LagKM3} and our discussion above it is clear that below the electroweak scale we simply deal with a new massive vector particle that has a coupling to charged particles according to Eq.~\eqref{xcharge}. This particle can then be produced in scattering experiments with charged particles. The production process is similar to that of photons only that the particle is massive and the interaction is suppressed by a factor $\chi$.
One incarnation of this are so-called fixed-target experiments\cite{Bjorken:2009mm}. In these experiments a large number of charged particles (often electrons) with a fairly high energy is shot onto a block or a foil (typically of metal). In the interaction between the incident and the target particles $X$s are produced.
If the mass of $X$ is now larger than $2m_{e}$, $X$ can and will decay into electron-positron pairs which can then be detected. 
The produced electron-positron pairs have to be distinguished from those produced in ordinary electromagnetic interactions. This can be done in two ways.
First the invariant mass distribution of the electron-positron pairs produced via an $X$ has a clear peak at the mass $m_{X}$. Second, for very low values of $\chi$ the produced $X$ is very long lived and we therefore have very displaced vertices (sometimes displaced by 100s of meters).
Bounds from this type of experiment\cite{Bjorken:2009mm,Merkel:2011kl,Abrahamyan:2011tg,Andreas:2012kx} are typically in the MeV to GeV region as shown in Fig.~\ref{exfig}.

Similarly scattering can be done at very high energy colliders such as the LHC where one typically looks for peaks in the invariant mass distribution\cite{Frandsen:2012hc,Jaeckel:2012zr}. The only difference is that in this region the mixing is with the hypercharge and we have an effective charge of $X$, 
\begin{equation}
\label{charges}
Q_{X}=\chi_{_Y} g^{\prime}\left[\frac{\gamma}{\tan^{2}(\theta_{W})}T^{3}-(1+\gamma)Q_{Y}\right],\quad {\rm where}\quad
\gamma=\tan^{2}(\theta_{W})\frac{m^{2}_{W}}{m^{2}_{X}-m^{2}_{Z}}.
\end{equation} 
The limits scaled to the electromagnetic mixing parameter are shown at the very high mass end of Fig.~\ref{exfig}.

Instead of real particle production we can also look at loop-effects such as (g-2) of the electron and the muon. The tight constraints in the MeV-GeV range from
precision measurements of these quantities are shown in Fig.~\ref{exfig}, labelled ``$a_{e,\mu}$''. 
As is well known, the muon (g-2) has a slight deviation from the SM expectation. HPs in a suitable range, shown red in Fig.~\ref{prospects} can 
fit this\cite{Pospelov:2008zw}. It should be noted that over the last couple of years this region has shrunk considerably due to a renewed experimental and theoretical effort. Note, however, that most of these limits hold only when the dominant decay of the HP is into SM particles. If there are additional ``hidden sector'' decay modes, fixed target and similar constraints will be relaxed, while it is still possible to fit (g-2).

\subsection{Option (2): Photon--hidden photon oscillations}

Let us now turn to the second option of dealing with the kinetic mixing term. Before we proceed let us, however, stress again that physical results are, of course absolutely independent of which picture we choose.

Inserting the shift $X^{\mu}\rightarrow X^{\mu}-\chi A^{\mu}$ into Eq.~\eqref{LagKM2} we obtain,
\begin{equation}
\label{LagKM4}
\mathcal{L} \supset -\frac{1}{4} F_{\mu\nu}F^{\mu\nu}- \frac{1}{4} X_{\mu \nu} X^{\mu \nu}
+ \frac{m_{X}^2}{2}\left( X_{\mu} X^{\mu} -2\chi X_{\mu}A^{\mu}+\chi^2 A_{\mu}A^{\mu}\right)+j_{\mu}A^{\mu}.
\end{equation}

With this shift we have again succeeded in removing the kinetic mixing term. Moreover, charged particles are still charged only under $A$ in contrast to the previous subsection. However, we now have a non-diagonal mass term, mixing $X$ and $A$.
This mass term now leads to $A\leftrightarrow X$ oscillations, in complete analogy to the non-diagonal mass matrices of neutrinos that lead to neutrino flavor oscillations.  Indeed this analogy goes even further. In the basis used in this section we are in the interaction eigenbasis and have oscillations. In the previous section we were in the mass eigenbasis and had non-trivial couplings to charged matter, but no 
oscillations\footnote{In the mass eigenbasis the ``oscillations'' appear as follows. Both mass eigenmodes couple to charged particles. Accordingly to know the effect on a charged particle we have to add the amplitudes of both eigenmodes multiplied with their respective couplings. This sum exhibits
the oscillatory behavior because the two eigenmodes with different masses acquire different phases when moving in space (similar in frequency).}.

The interaction basis we are discussing in this section is particularly convenient when the interactions with matter are in some sense ``strong''.
An example are the so-called light-shining-through-walls experiments (LSW). In these experiments a laser is shone onto an opaque wall and one looks for light ``coming out of the wall''. The idea is that an initial photon oscillates into a hidden photon, which traverses the wall unimpeded and after the wall
oscillates back into a photon that is subsequently detected. For this it is convenient to use the interaction basis, as it is only the $A$ component that interacts with the wall (and the laser and the detector) and is completely suppressed (this is the ``strong'' effect caused by the large amount of interacting particles in the wall). Current limits\cite{Ehret:2010dq} from ``LSW'' are shown in Fig.~\ref{exfig}.  

The centre of stars contains an extremely high number of photons which can oscillate into HPs and then leave the star (the rest of the star is basically a thick wall). This is a very efficient way for stars to loose energy. In this kind of environment the oscillations are modified by the presence of a plasma allowing for resonances (similar to the MSW effect for neutrinos) but also allowing the longitudinal modes to contribute. 
Limits on an extra energy loss for the sun and horizontal branch stars give extremely tight constraints\cite{Redondo:2008aa,An:2013vn} (cf. Fig~\ref{exfig}).

Similarly production of HPs in the early universe plasma either before the CMB release\cite{Redondo:2009cr} (region ``thermal cosmology'') or 
after\cite{Jaeckel:2008fi,Mirizzi:2009nx} (region ``CMB'') benefits from resonances and leads to a very sensitive test.

\subsection{Have it your way}
As stressed before observables are independent of the chosen picture. But in the above examples the description may be more transparent in one picture.
In others no clear preference is obvious.

For example the regions ``$BR(\pi^{0}\to \gamma e^{+}e^{-})$''\cite{Gninenko:2013qy} and ``$e^{+}e^{-}\to \gamma \mu^{+}\mu^{-}$'' arise from decays of pions and decay of $\Upsilon$ resonances\cite{Echenard:2012oq,Reece:2009ij,Essig:2010xa,Hook:2011fv}, respectively. 
These mesons can decay into two photons. In presence of a kinetically mixed hidden photon we can now have one photon oscillate 
into a HP which subsequently decays into an electron-positron pair with an invariant mass $\approx m_{X}$. Alternatively one can take picture (1) and imagine that the meson couples (via the charged quarks) to a photon and a hidden photon.

For the electroweak precision, ``EW'', constraints\cite{Hook:2011fv} arising most notably from the linewidth and lineshape of the Z resonance one has to take into account the full electroweak symmetry breaking and mixing effects. 

\section{Searching hidden photons: The future}
\begin{figure}[t]
    \begin{center}
         {\includegraphics[scale=0.55]{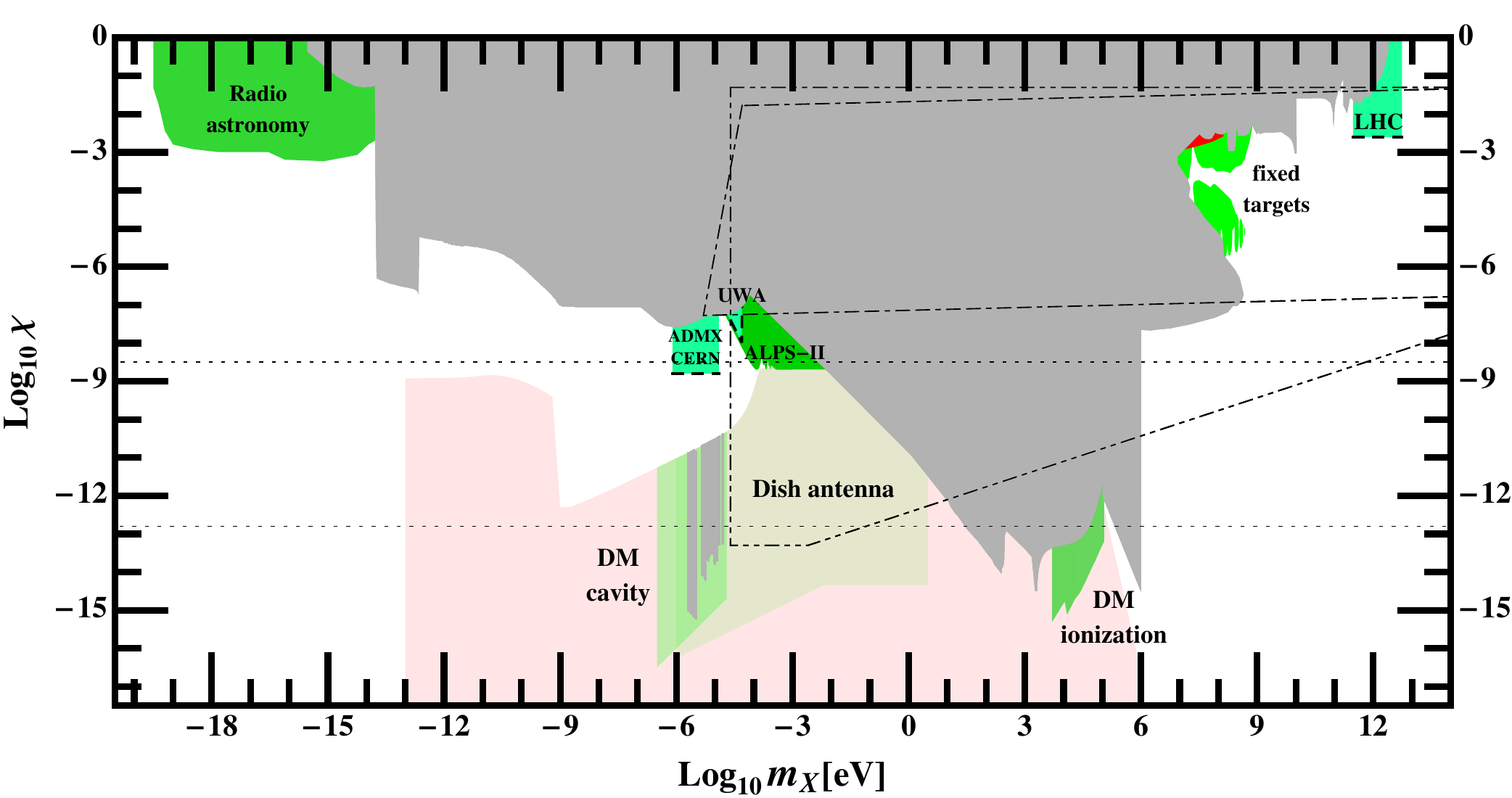}}
        \caption{\it Prospects for near future searches for hidden photons (shaded green). The regions labeled, ADMX, CERN and UWA are LSW experiments in the microwave range (cf.\cite{Hewett:2012ys}), ALPS-II will operate in the optical regime\cite{Bahre:2013qf}. Radio astronomy searches for a frequency dependence of distant sources could provide sensitivity at very small masses\cite{Lobanov:2012pt}. The dish antenna region gives the prospects for a broadband search for dark matter HP according to\cite{Horns:2012ve}. At higher masses we have fixed target 
experiments\cite{Hewett:2012ys,Merkel:2012qa} and the LHC as well as a detection of ionization signals, caused by HP DM in direct detection experiments for WIMP DM\cite{Pospelov:2008jk}. The small red region at high masses corresponds to the $(g-2)_{\mu}$ hint\cite{Pospelov:2008zw}.}
\label{prospects}
    \end{center}
\end{figure}

In the previous section we have reviewed existing constraints on hidden photons,
spanning a wide range of masses from $10^{-15}\,{\rm eV}$ to more than $10^{12}\,{\rm eV}$.
Yet, the search is far from over. 
In Fig.~\ref{prospects} we compare the existing constraints (grey areas) with interesting target regions from models of string theory (regions enclosed by dashed, dotted and dashed-dotted lines)\cite{Goodsell:2009xc,Cicoli:2011yh}. The hint from (g-2)$_{\mu}$\cite{Pospelov:2008zw} (red) and the region where
hidden photons can be dark matter (light red)\cite{Arias:2012ly} (and next section).
It is obvious that there are large regions that still need to be explored. In this section we look at the near future experimental prospects (shown in 
different shades of green).

\subsection{Low energy probes}
A clear step forward in the ${\rm meV}$ mass range will be the next generation of LSW experiments,
most notably ALPS-II\cite{Bahre:2013qf} (cf. the corresponding region in Fig.~\ref{prospects}) and REAPR. 
These experiments will feature a resonant regeneration scheme\cite{Hoogeveen:1990vq,Sikivie:2007qm}. In such a setup the light is reflected back and forth in an optical cavity on the production side (this was already pioneered by ALPS-I\cite{Ehret:2009sq}) as well as on the regeneration side. This enhances the signal by a factor $N_{\rm prod}N_{\rm reg}$ where $N_{\rm prod}$ and $N_{\rm reg}$ are the number of passes on the production and regeneration side, respectively.
As $\sim100000$ passes seem feasible this yields an enormous gain in sensitivity. Moreover, ALPS-II will be significantly longer, allowing to probe also smaller masses.

At even smaller masses in the $\mu{\rm eV}$ region, LSW with ``light'' in the microwave range\cite{Hoogeveen:1992uk,Jaeckel:2007ch} is very promising. Here, resonant regeneration is easier to achieve and first experiments have already been performed\cite{Povey:2010hs,Wagner:2010mi,Betz:2012dz}.
Such experiments also benefit from a potentially enormous number of passes of up to $10^{11}$ in the microwave cavities and very sensitive detection techniques\cite{Caspers:2009cj}. This allows for very interesting sensitivity as shown by the regions ``ADMX CERN'' and ``UWA'' in Fig.~\ref{prospects}.

At even smaller masses it may also be possible to probe photon-hidden photon oscillations by observing the frequency dependence of well 
modelled radio signals\cite{Lobanov:2012pt}. The frequency dependence of the oscillation probability would lead to structures in the
frequency spectrum at a frequency scale $\Delta \omega \sim \omega^2/(m^{2}_{X} L_{\rm source})$ (where $L_{\rm source}$ is the distance to the source) and amplitude $\sim \chi^2$. This will extend sensitivity to extremely small masses as shown by the ``Radio astronomy'' region in Fig.~\ref{prospects}.   

In the sub-eV region one can also search for HP dark matter with extremely high sensitivity as we will see in the next section.

\subsection{Intermediate energy probes}
In the MeV to GeV region the next few years will bring new and improved fixed target experiments. In particular the A1 collaboration with the MAMI and MESA accelerators at Mainz\cite{Merkel:2012qa} as well as APEX\cite{Essig:2010xa}, Darklight\cite{Freytsis:2010mi} and HPS\cite{HPS} at Jefferson Lab will use electrons on various targets whereas the VEPP-3\cite{Wojtsekhowski:2009ff} at SLAC intends to use positrons.
The region projected to be tested by these experiments is shown as the green ``fixed target'' region in Fig.~\ref{prospects}. It encompasses all of the (g-2)$_{\mu}$ region. In addition proton fixed target experiments promise additional information\cite{Batell:2009di}.

\subsection{High energy probes}
The LHC will continue to improve the limits on hidden photons in the $100\,{\rm GeV}$ to multi TeV range over the next years.
The first step in the improvement will be the increase in the beam energy from 8 TeV to 13 TeV hopefully even 14 TeV. This will approximately double 
the mass reach for hidden photons. Increased integrated luminosity collected over the next few years will importantly improve the sensitivity to smaller cross sections and therefore smaller kinetic mixings. 
Further improvements towards somewhat smaller masses may also come from analyses that include lower energy electrons and muons.
An indication of the tested region is the green ``LHC'' region in Fig.~\ref{prospects}.

\section{Hidden photon dark matter}
In contrast to common belief very light particles can indeed be good dark matter candidates. The crucial aspect is that {\emph{very light dark matter
particles need to be produced non-thermally.}} The most famous example is the so-called misalignment mechanism 
for the axion\cite{Preskill:1982cy,Abbott:1982af,Dine:1982ah}.
Its essential features are as follows. 
In the early Universe the field starts with a non-vanishing expectation value. Because the Hubble constant is much larger than the mass of the field in question, the field is basically stuck at this initial value. This is also the reason why it is not unreasonable to assume a non-vanishing initial value; there is simply no time to relax to the minimum of the potential. Moreover, after inflation the field is smoothed out and basically has the same field value everywhere in space. 
Once the Hubble value drops below the mass of the field, the field starts to oscillate and one can show that these oscillations behave like non-relativistic matter. This becomes plausible by remembering that the momentum of a particle is essentially the spatial derivative of the field, and a (nearly) homogeneous field configuration therefore corresponds to particles with extremely low momentum.

As argued recently\cite{Nelson:2011sf,Arias:2012ly} a similar mechanism works also for hidden photons. 
For the evolution in the early Universe the Hubble constant is larger than the mass. It is therefore reasonable
to include additional interactions with gravity,
\begin{equation}
\label{hidden}
\mathcal L= -\frac{1}4X_{\mu\nu}X^{\mu\nu}+\frac{m_{X}^2}2 X_\mu X^\mu + \frac{\kappa}{12}R X_{\mu}X^{\mu}+{\mathcal L}_{\rm kinetic\,\, mix}+{\mathcal L}_{\rm SM},
\end{equation}
where $R$ is the Ricci scalar\footnote{{We use a 
coordinates such that $ds^2=dt^2-a^{2}(t)dx^{2}_{i}$, i.e. the metric is $g_{\mu\nu}={\rm diag}(1,-a^{2}(t),-a^{2}(t),-a^{2}(t))$. Moreover, the
gravitational part of the Lagrangian is ${\cal L}_{\rm GR}=-R/(16\pi G_{N})$.}} and the last two terms indicate the kinetic mixing and the rest of the Standard Model.    

As already mentioned in Sect.~\ref{sec:hid}, in general, the HP mass might result from a Higgs or a St\"uckelberg mechanisms. In the first case, we have to worry when the phase transition happens and also take into account the effects of the Higgs field.
As in\cite{Nelson:2011sf,Arias:2012ly}, we focus therefore on the St\"uckelberg case, which occurs naturally 
in large volume string compactifications\cite{Burgess:2008ri,Goodsell:2009xc,Cicoli:2011yh}. 
In this case, there is no phase transition.

Let us briefly comment on the equations of motion and in particular determine the behavior of the energy density in the expanding universe.
Let us see if we can find the hallmark property of non-relativistic ``cold'' dark matter: the energy dilutes like the volume, i.e. $\rho\sim 1/(a(t))^{3}$ where $a(t)$ is the scale factor. 

For simplicity let us focus on the homogeneous solution, $\partial_{i}X_{\mu}=0$.
The equation of motion then enforces $X_{0}=0$.
As explained in\cite{Golovnev:2008cf} the invariant $X^{\mu}X_{\mu}=-1/a^{2}(t) X_{i}X_{i}$ is a coordinate independent measure for the size of the vector and it is convenient to introduce $\bar{X}_{i}=X_{i}/a(t)$. 
Using this the equation of motion is,
\begin{equation}
\label{vecevolution}
\ddot{\bar{X}}_{i}+3 H\dot{\bar{X}}_{i}+\left(m^2_{X}+(1-\kappa)(\dot{H}+2H^2)\right)\bar{X}_{i}=0.
\end{equation}
The energy density is
\begin{equation}
\label{vecdensity}
\rho(t)=T^{0}_{0}=\frac{1}{2}\left(\dot{\bar{X}}_{i}\dot{\bar{X}}_{i}+m^{2}_{X}\bar{X}_{i}\bar{X}_{i}+(1-\kappa)H^2 \bar{X}_{i}\bar{X}_{i}+2(1-\kappa) H\dot{\bar{X}}_{i}\bar{X}_{i}\right).
\end{equation}

For $H\ll m_{X}$ and $\dot{H}\ll m^{2}_{X}$ Eq.~\eqref{vecevolution} is that of a weakly damped harmonic oscillator.
One can easily see that the amplitude oscillates with frequency $m_{X}$ and it is damped by a factor $\exp(-3\int H dt/2)\sim 1/a^{3/2}(t)$.
Inserting into Eq.~\eqref{vecdensity} we then find that the energy density indeed dilutes like the volume $\sim 1/a^{3}(t)$.
Taking a different point of view: In this limit both expressions~\eqref{vecevolution} and \eqref{vecdensity} reduce
to same form as the equations of motion and the energy momentum tensor of a scalar field (independent of the value of $\kappa$) and for such a 
field we already know (e.g. from the case of axions) that this behaves as non-relativistic matter.
For $\kappa=1$ this equivalence holds true exactly even in the very early Universe where $H\gtrsim m_{X}$ and/or $\dot{H}\gtrsim m^{2}_{X}$.

Now we have seen that an initial field value for hidden photons can nicely behave like cold dark matter. However, there are several questions remaining: 1) How do we get the right abundance. 2) Does it survive? 3) Does it change observation?
For a detailed discussion see\cite{Arias:2012ly} but let us briefly summarize the answers. 1) The abundance is proportional to the initial field value squared. Therefore getting the right abundance requires some amount of fine-tuning or anthropic arguments. For 2) and 3) we have to make sure
that the HPs forming do not decay, or spoil observations such as, e.g. the CMB. This leads to upper limits on the kinetic mixing parameter.
Heavier HPs can decay into electron-positron pairs via the kinetic mixing interaction. This is quite fast and therefore basically\footnote{Unless one considers\cite{Chen:2008yi} incredibly small kinetic mixings $\chi\lesssim 10^{-26}$.} 
 rules out HP dark matter with masses $m_{X} \gtrsim {\rm MeV}$.
At lower masses HPs can decay into three photons, but this decay is fairly suppressed by a by a small algebraic factor as well as phase space. Therefore it provides a constraint only at masses just below an MeV. 
At even smaller masses other evaporation mechanisms originating from photon-HP mixing in the plasma of the early universe and scattering with the electrons in the plasma dominate. This both reduces the condensate as well as transferring energy to the ordinary electrons and producing extra photons. This can lead to observable effects such as distortions in the CMB or a changed number of effective neutrinos. Although all these effects combined set fairly strong constraints, a large and interesting area of viable HP dark matter remains. This is shown as the light red area in Figs.~\ref{exfig} and \ref{prospects}.

\subsection{Detecting hidden photon dark matter}
Having a viable dark matter candidate it is now desirable to also have ways to detect it.

First of all, let us note that due to the small mass of HP dark matter particles, conventional direct detection methods designed for weakly interacting massive particles (WIMPs) based on nuclear or electronic recoils do not work, the recoil energy is just too small.

Nevertheless, we have a plentiful source of HPs and we have already seen previously that those have a natural tendency to convert into ordinary photons that can be detected.
Therefore, in principle we do not need to do anything but simply set up a detector for photons and wait. As dark matter HPs are very slow their energy is essentially given by their mass and therefore we expect a nicely peaked signal of photons with energy/frequency corresponding to the HP mass.

Unfortunately things are not quite as easy since the rate of produced photons is quite small.
Therefore we need incredibly good photon receivers as well as shielding of background noise. Moreover, it is desirable to have techniques to enhance the signal. Two options are currently being pursued. The first is based on the haloscope technique originally 
proposed for axions\cite{Sikivie:1983ip} but it can also be used for detecting hidden photons\cite{Arias:2012ly}.
The idea is to enhance the conversion of HPs into photons by employing a suitable cavity resonant with the HP mass. This is analog
to the resonant regeneration scheme discussed above for LSW experiments. The output power is then enhanced by the $Q$-factor of the 
cavity (which is proportional to the number of reflections),
\begin{equation}
P_{\rm out}\sim \chi^{2}m_{X}\rho_{\rm HP} Q V {\mathcal G},
\end{equation}
where $\rho_{\rm HP}$ is the density of dark matter hidden photons, $V$ the volume of the cavity and ${\mathcal G}$ a geometrical factor encoding properties of the cavity as well as the dark matter configuration.
This technique works particularly well in the microwave regime where good cavities as well as excellent detectors exist. This technique has 
already been employed\cite{DePanfilis:1987dk,Wuensch:1989sa,Hagmann:1990tj,Asztalos:2001tf,Asztalos:2009yp} and further improvements are 
underway~\cite{Baker:2012kx,Asztalos:2011bm,Heilman:2010zz}. The currently excluded region is the ``haloscope region'' shaded in 
grey in Figs.~\ref{exfig} and future prospects are indicated as the ``DM cavity'' region in Fig.~\ref{prospects}.

The cavity technique is sensitive to extremely small couplings but it has two drawbacks. As it relies on the resonant enhancement
it requires a slow scan through masses, where each measurement is only sensitive to a small region of masses and then the cavity has to be tuned to a new frequency and another measurement has to be performed. In essence one needs to do a slow and time-consuming scan through the masses. 
The second issue is that the output power is proportional to the volume of the cavity. With increasing frequency the volume of the cavity goes down, or one has to operate the cavity in a higher mode which often leads to smaller geometrical factors.

Accordingly it would be nice to have a broadband technique without a ``volume suppression''. One option is to use a ``dish antenna''\cite{Horns:2012ve}. Here the basic idea is as follows. On charged matter the HP dark matter field essentially acts like a small oscillating ordinary electric field.
The electrons in a conducting (reflecting) surface then start to oscillate in this field, emitting ordinary photons. One can check that for very slow HPs the produced electromagnetic radiation is emitted perpendicular to the surface. Using a suitable (spherical) surface all radiation from the whole surface is concentrated in a point, the centre, where it can be detected. It is easy to understand that the concentrated power in the centre is proportional to the area $A_{\rm dish}$ of the ``antenna'',
\begin{equation}
P_{\rm centre}\sim \chi^2\rho_{\rm HP} A_{\rm dish}.
\end{equation}
This technique can be used in the radio frequency range but also promises good sensitivity at much higher frequencies in the optical regime.
This is indicated as the very light green ``Dish antenna'' region in Fig.~\ref{prospects}.

Finally, experiments for direct detection of WIMPs may also be used\cite{Pospelov:2008jk}. Dark matter HPs with masses in the multi keV may be completely absorbed. This produces an ionization signal that can be detected. A rough estimate for a future sensitivity is shown as ``DM ionization'' in Fig.~\ref{prospects} (we have assumed that the direct detection experiment is sensitive to an ionization signal from an $0.1\,{\rm fb}$ cross section at WIMP masses of $\sim 100\,{\rm GeV}$; other assumptions as in\cite{Pospelov:2008jk}).

\section{Conclusions}
New forces are a hallmark feature of many models of physics beyond the Standard Model. A very simple but also very well motivated example
is an extra U(1) gauge force kinetically mixed with the electromagnetic or hypercharge U(1) of the Standard Model. 
In this note we have given overview over past and future searches for such a new force and the corresponding gauge boson, often called hidden photon, heavy photon, dark photon or $A^{\prime}$.

The simple example of a hidden photon nicely demonstrates that there are (at least) two directions to explore. The hidden photon could be 
very light and very weakly coupled, corresponding to a very weak but long range force. Alternatively it could be heavy in which case the interaction is very short range (and weak in this sense).  

Searches for hidden photons span a huge range of masses and energies from $10^{-20}\,{\rm eV}$ to multi-TeV. Exploring this vast range requires
to exploit the complementarity between very different experimental techniques, from low energy ultrahigh precision experiments to high energy colliders. This complementarity can be nicely seen in Fig.~\ref{exfig} where the different energy scales probed by a variety of experiments nicely fit together to cover a large range in masses and couplings.

While significant progress has been made over the past few years, huge regions remain to be explored.  
Interesting regions are suggested by models of fundamental theories (e.g., string theory) but also by the intriguing possibility that
hidden photons form all or part of dark matter. 
It is therefore very exciting that many experiments, exploring different energy/mass ranges and using a variety of techniques 
are underway, in construction or in planning, promising new insights into fundamental physics and even the potential for discovering a new force.

\section{Acknowledgements}
JJ would like to thank Javier Redondo and Andreas Ringwald for joyful long-term collaboration on the subject.

\end{document}